\documentclass[showpacs,eqsecnum,aps,nofootinbib,preprintnumbers,floatfix]{revtex4}

\usepackage{epsfig,citesort}

\def  \ktpnn {$K_L \to \pi^0 \nu  \bar{\nu}~$} 

\def  \bcen  {\begin{center}}
\def  \ecen  {\end{center}}

\def  \etal  {{\sl et.al.},}
\def  \beq   {\begin{equation}}
\def  \eeq   {\end{equation}}
\def  \beqa  {\begin{eqnarray}}
\def  \eeqa  {\end{eqnarray}}
\def  \bfig  {\begin{figure}}
\def  \efig  {\end{figure}}

\begin{document}

\renewcommand{\thefootnote}{\fnsymbol{footnote}}

\title{\boldmath $K_L \to \pi^0 \nu \bar{\nu}~$ in Little Higgs model} 

\author{S. Rai Choudhury}
  \email{src@physics.du.ac.in}
  \affiliation{ Department of Physics \& Astrophysics, \\
   University of Delhi, Delhi - 110 007, India.}
\author{Naveen Gaur}
  \email{naveen@physics.du.ac.in}
  \affiliation{ Department of Physics \& Astrophysics, \\
   University of Delhi, Delhi - 110 007, India.}
\author{G. C. Joshi}
  \email{joshi@tauon.ph.unimelb.edu.au}
   \affiliation{School of Physics, \\ 
    University of Melbourne, Victoria 3010 , Australia.}
\author{B. H. J. McKellar}
   \email{b.mckellar@physics.unimelb.edu.au}
   \affiliation{School of Physics, \\ 
    University of Melbourne, Victoria 3010 , Australia.}

\begin{abstract}
\noindent Little Higgs models provides us a new solution of the {\it
hierarchy problem} of the SM. The problem is associated with the large
difference in scales of SM and plank scale. One of the main features 
of these models is the existence of a vector like top quark. 
The CP violating \ktpnn process n the SM model is dominated by top
quark penguin \& box graphs.  
We examine this process in the Little Higgs model where the
top quark sector has significant differences from the Standard Model. 
\end{abstract}


\maketitle

\setcounter{footnote}{0}

\section{Introduction \label{section:1}}

The Standard Model (SM) of particle physics is a extremely successful
theory. Electroweak precision tests have probed SM at quantum
level and have confirmed predictions of SM. The symmetry breaking
sector (Higgs sector) of the SM has also been investigated by these
precision measurements and it indicates the existence of weakly
coupled Higgs sector and light Higgs boson whose mass is $m_H \le 200$
GeV. This existence of weakly coupled Higgs sector creates what is
known as {\it hierarchy problem}. This problem is related to the
origin and radiative stability of two widely different mass scales
namely the Electroweak (EW) scale and Plank scale. To solve this
problem the Higgs sector has to be fine tuned from EW scale to plank
scale. There 
have been many suggestions to avoid this problem. In one of the
attractive solutions namely Supersymmetry (SUSY) the quadratic
divergences in Higgs mass are canceled between fermionic and bosonic
loops provided the SUSY breaking scale is near TeV. Extra dimensions
theories use the geometry of higher dimensional space-time to address
the problem. 

\par An alternative approach to solve this hierarchy problem has been 
recently considered in generically called {\it ``Little Higgs model''}
\cite{Arkani-Hamed:2001nc,Han:2003wu}. The basic idea in these models
is to realize Higgs boson   
doublets (and other scalars) as Goldstone modes in a globally SU(5)
symmetric theory, spontaneously broken at a scale $f$ in the TeV range
much higher than the vev of the SM Higgs ($v$)
\cite{Han:2003wu,Schmaltz:2002wx}. In the simplest 
version of the theory known as {\it Littlest Higgs} model
\cite{Han:2003wu} , the effective theory at low energies involves many
more particles in addition to the SM particles. Thus apart from SM
spectrum there are charged heavy vector bosons ($W_H$), neutral
heavy vector boson ($Z_H$), heavy photon ($A_H$), triplet of charged
Higgs ($\Phi^{++}, \Phi^+, \Phi^0$) and a heavy top quark ($T$). The
masses of these heavy particles are expected in the TeV region
\cite{Han:2003wu,Schmaltz:2002wx}. All these particles are expected to
provide $O(v^2/f^2)$ corrections to all Flavor Changing Neutral
Current (FCNC) amplitudes which are generated through loops. In
addition because of mixing of the SM t-quark and its heavier
counterpart T, we expect $O(v^2/f^2)$ violation of the CKM unitarity
relation. A host of processes
\cite{Han:2003wu,Han:2003gf,Choudhury:2004bh,Chen:2003fm,Huo:2003vd}
have been evaluated in this model providing constraints on the vast
parameter space of the LH, which will be useful in experimental search
for the validity of the model.

\par The CP violating FCNC process \ktpnn with an expected SM
branching ratio of $\sim 10^{-10}$ is another process which is of
special interest from theoretical viewpoint. The importance of this
process in SM is because of two fold reasons firstly its proceeds
through direct CP violation and secondly is totally dominated by short
distance top-quark loops and charm quark plays no role in it
\cite{Fleischer:2002ys}. Because of these reasons this process is
believed to be the most ideal one for extracting out the CP violating
Wolfenstein parameter $\eta$. Further, in view of the large mass of
the top quark, QCD corrections are small and calculable in
perturbation theory with the result that one expected the basic SM
graphs to reproduce the amplitude quite accurately
\cite{Fleischer:2002ys}. The article by Buchalla \& Buras
\cite{Buchalla:1998ba} reviews and 
updates the SM predictions for this process with a more complete set
of references. A recent review by Isidori \cite{Isidori:2003ij} covers
the same ground with a summary of new physics possibilities. 
This process has been extensively studied in literature for finding
out signatures in new physics
\cite{Deshpande:2004xc,Fleischer:2002ys,Buras:2004uu,Buras:2004ub,He:2004it}.

\par The LH models has substantial modification to top-quark loops,
both in terms of Unitarity relation violation and extra loops arising
out of replacing t by T-quark
\cite{Choudhury:2004bh,Huo:2003vd}. As its known that FCNC processes
vanishes if we have a unitarity of CKM and complete horizontal
symmetry of the masses of the quarks {\sl i.e.}, masses of all the
quarks are same. In SM although CKM is unitary but we the large
t-quark mass breaks the horizontal symmetry of quark masses which
results in low rates of FCNC processes. Many of these low energy FCNC
processes crucially depends on the top quark mass. Another
theoretically very important process which depends on top quark mass
in \ktpnn because of the reasons mentioned above. 
Experimentally, at 90\% CL, we have an indirect limit for the
Branching ratio for the process $Br(K_L \to \pi^0 \nu \bar{\nu}) < 1.7
\times 10^{-9}$ \cite{Isidori:2003ij} which in principle can be
achieved at SM  level in the future dedicated experiments
\cite{Isidori:2003ij}. As the LH models predicts a new top quark with
mass in TeV range so it is worthwhile to test the effects which LH
type models have on this decay.   

\par This paper is organized as follows : in Section \ref{section:2}
we will present the effective Hamiltonian for the process in SM. In
sections \ref{section:3} we will present the results of the
corrections due to $T$ quark. In Section \ref{section:4} results of
corrections due to extra scalars is given and finally in
Section \ref{section:5} we will present the results of the correction 
due to heavy photon $A_H$. In the last section \ref{section:6} we will
conclude with numerical analysis of our results and discussion.  


\section{Effective Hamiltonian \label{section:2}} 

\par The basic quark level graphs in SM responsible for \ktpnn are
shown in Figures
\ref{dia:1},\ref{dia:2},\ref{dia:3}. The
effective Hamiltonian for the process can be 
written as : 
\beq
{\cal H}_{eff} =  \frac{G_F}{\sqrt{2}} \frac{\alpha}{2 \pi
Sin^2\theta_W} (V_{ts}^* V_{td}) (\bar{s} d)_{V-A} (\bar{\nu}
\nu)_{V-A} X(x_t) + h.c. 
\label{sec2:eq:1}
\eeq
where $x_t = \left(\frac{m_t^2}{m_W^2}\right)$ and the function $X$ in
SM has been worked out to $O(\alpha_s)$ in \cite{Fleischer:2002ys}. We
will evaluate the additional contributions to $X(m_t)$ given by
particles of LH model. Since the
result will be the corrections to the SM value, we do not calculate
the $O(\alpha_s)$ corrections to this. For our calculations we have
used unitary gauge, unlike the original SM calculation
\cite{Inami:1980fz} but check that the total SM contribution matches
with the results 
given by Buchalla \& Buras \cite{Buchalla:1995vs}. Further we retain
terms upto order $O(v^2/f^2)$ and consequently drop all terms 
in any diagram which is of order higher than ${\cal O}(v^2/f^2)$. We 
also drop terms independent of the internal quark mass since the CKM
unitarity is 
valid upto this order when the T-quark is also considered 
in LH. For the results of the calculation of the individual diagrams
we drop the divergent parts as we have checked that
these divergences cancel when all the diagrams are added in LH model.

We have used Feynman rules given in Han \etal \cite{Han:2003wu}.
Before presenting our results first we will define our
convention. In our convention the indices $i,k,l,a$ have values
$1,2$. In our convention  
$$
\begin{array}{c c l l c }
     &        &                 &              &  Masses  \\
t_i  &: i = 1 & \Longrightarrow &  t ~~~~({\rm SM ~ top}) & m_1 = m_t
\\ 
     &: i = 2 & \Longrightarrow &  T ~~~~({\rm extra ~ vector ~ like ~
top ~ quark})  & m_2 = m_T \\
W_i  &: i = 1 & \Longrightarrow &  W_L  & M_1 = m_{W_L}
\\ 
     &: i = 2 & \Longrightarrow &  W_H  & M_2 = m_{W_H} \\
Z_i  &: i = 1 & \Longrightarrow &  Z_L  & M_{Z_1} = m_{Z_L}
\\ 
     &: i = 2 & \Longrightarrow &  Z_H  & M_{Z_2} = m_{Z_H} \\
\end{array}
$$
The vertices's involved in our calculations are summarized in Appendix
\ref{appendix:b}. 

\par In LH model the SM function $X(x_t)$ given in
eqn.(\ref{sec2:eq:1}) gets additional contributions from new graphs
having LH particles. The new value of the function in LH model becomes
\footnote{in writing this we have only retained terms of order
${\cal O}(v^2/f^2)$ }: 
\beqa
X^{LH} &=& X^{SM}(x_t) + X_{Z_L,Z_H}(T W_L, T W_H, t W_H, t W_L) 
 + X_{A_H}(t W_L) + X_{Z_{L,H}} (\Phi t, \Phi T) \nonumber \\
&& + ~ X^{Box}(W_k W_l t_i ~~{\rm except}~ k = l = i = 1) 
\label{eq:sec2:2}
\eeqa
where $X^{SM}$ represents the SM contribution, the other terms
represents the extra contribution in LH and are defined in sections
\ref{section:3}, \ref{section:4}, \ref{section:5}. 


\section{Corrections due to extra SM top and T quarks, W and Z bosons 
\label{section:3}} 

The Feynman diagrams corresponding to the contribution of $t, T$
quarks, $W_{L,H}$ and $Z_{L,H}$ are given in
Figures \ref{dia:1}(a)(b), Fig.\ref{dia:3}(a)(d) and Fig.\ref{dia:2} 
. We are not including the effects of scalars ($\Phi$) and heavy
photon ($A_H$), these results we will present in next section. In 
eqn(\ref{eq:sec2:2}) this contribution is represented by term
$X_{Z_L,Z_H}(T W_L, T W_H, t W_H)$ . This contribution comes from
feynman diagrams in figures  \ref{dia:1}(a)(b), Fig.\ref{dia:3}(a)(d)
and Fig.\ref{dia:2}. We rewrite this to form :   
\beqa
X_{Z_L,Z_H}(T W_L, T W_H, t W_H, t W_L) 
&=&  X_{Z_L}(T W_L, T W_H, t W_H) + X_{Z_H}(t W_L) \nonumber \\
&=& X^W(t) + X^t(t) + X^{SE}(t) 
\label{eq:sec3:1}
\eeqa
where $X^W$ and $X^t$ are the contributions of the penguin diagrams
where the neutral boson ($Z_{L,H}$) is emitted from charged vector
boson ($W_{L,H}$) and top quark respectively whereas $X^{SE}$
indicates the contribution of the self energy diagram. 
\bfig[ht]
\bcen
\epsfig{file=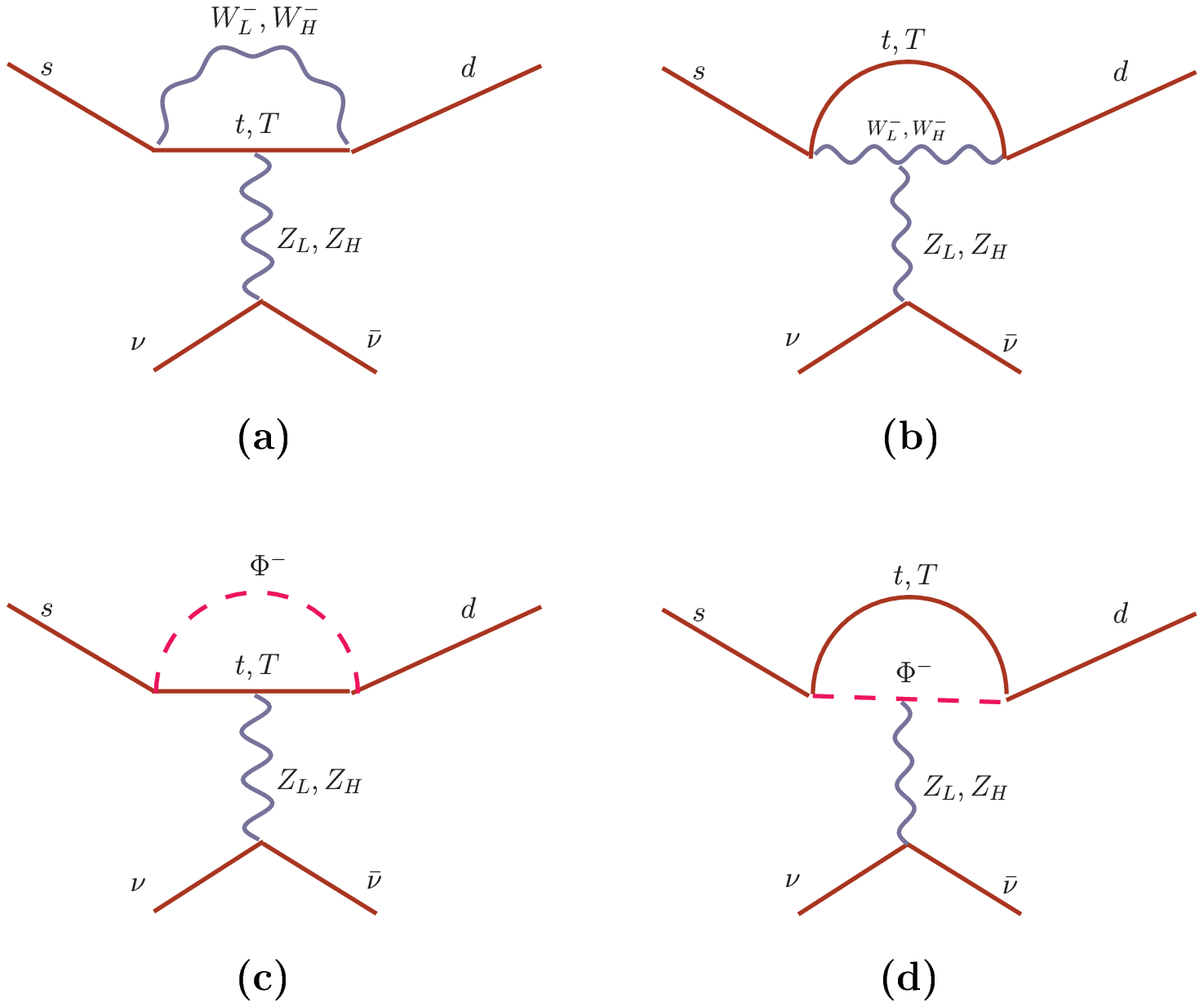,height=3in}
\caption{Penguin diagrams}
\label{dia:1}
\ecen
\efig
\bfig[ht]
\bcen
\epsfig{file=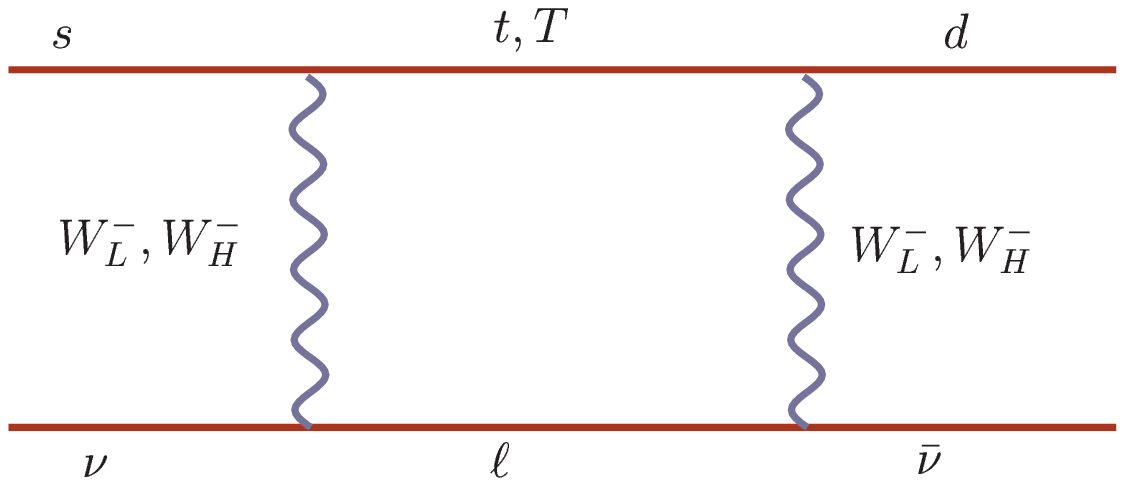,height=1.5in}
\caption{Box  diagram.}
\label{dia:2}
\ecen
\efig
\par The contribution of the diagram in Fig.\ref{dia:1}(b) where $Z$
is emitted from charge vector boson is :
\beqa
X^W(t) &=& \sum_{i,k,l,a} \left\{ \frac{8 \pi^2 M_{Z_l}^2
Cos^2\theta_w}{M_{Z_a}^2}  
g_{tsW}(i,k) g_{tdW}(i,l) g_{WWZ}(k,l,a) g_{\nu \nu
Z}(a) \right\}  
 \Bigg[ 3 m_i^2 F(m_i,M_k,M_l)   \nonumber  \\
&& - {3 \over 4} \frac{m_i^2 (M_k^2 + M_l^2)}{M_k^2 M_l^2} 
\left( F(M_k,M_l) + m_i^2 F(m_i,M_k,M_l) \right) 
+ \frac{1}{128 \pi^2} \frac{m_i^2 (M_k^2 + M_l^2)}{M_k^2 M_l^2} 
\Bigg]
\label{eq:sec3:2}
\eeqa
where the functions F's are defined in the appendix \ref{appendix:a}

\par Contribution of the diagram Fig.\ref{dia:1}(a) where $Z$ is
emitted from top quark line is :
\beqa
X^t(t) &=& - \sum_{i,j,k,a} \left\{ \frac{8 \pi^2 M_{Z_L}^2
Cos^2\theta_w}{M_{Z_a}^2} g_{tdW}(j,k) g_{tsW}(i,k) g_{\nu \nu Z}(a) 
\right\}        \nonumber \\
&&
\Bigg[ g^L_{ttZ}(i,j,k) 
  \left\{ F(m_i,m_j,M_k) 
     \left(M_k^2 + \frac{m_i^2 m_j^2}{M_k^2} - (m_i^2 + m_j^2) \right)
      + F(m_i,m_j) \left( 1 - \frac{(m_i^2 + m_j^2)}{M_k^2} \right) 
  \right\}   \nonumber \\ 
&&  + g^R_{ttZ}(i,j,k) m_i m_j \left\{ - {3 \over 2} F(m_i,m_j,M_k)
   + \frac{1}{2 M_k^2} F(m_i,m_j) 
     - \frac{1}{4 M_k^2} \frac{1}{16\pi^2}  \right\}
\Bigg] 
\label{eq:sec3:3}
\eeqa
\bfig[ht]
\bcen
\epsfig{file=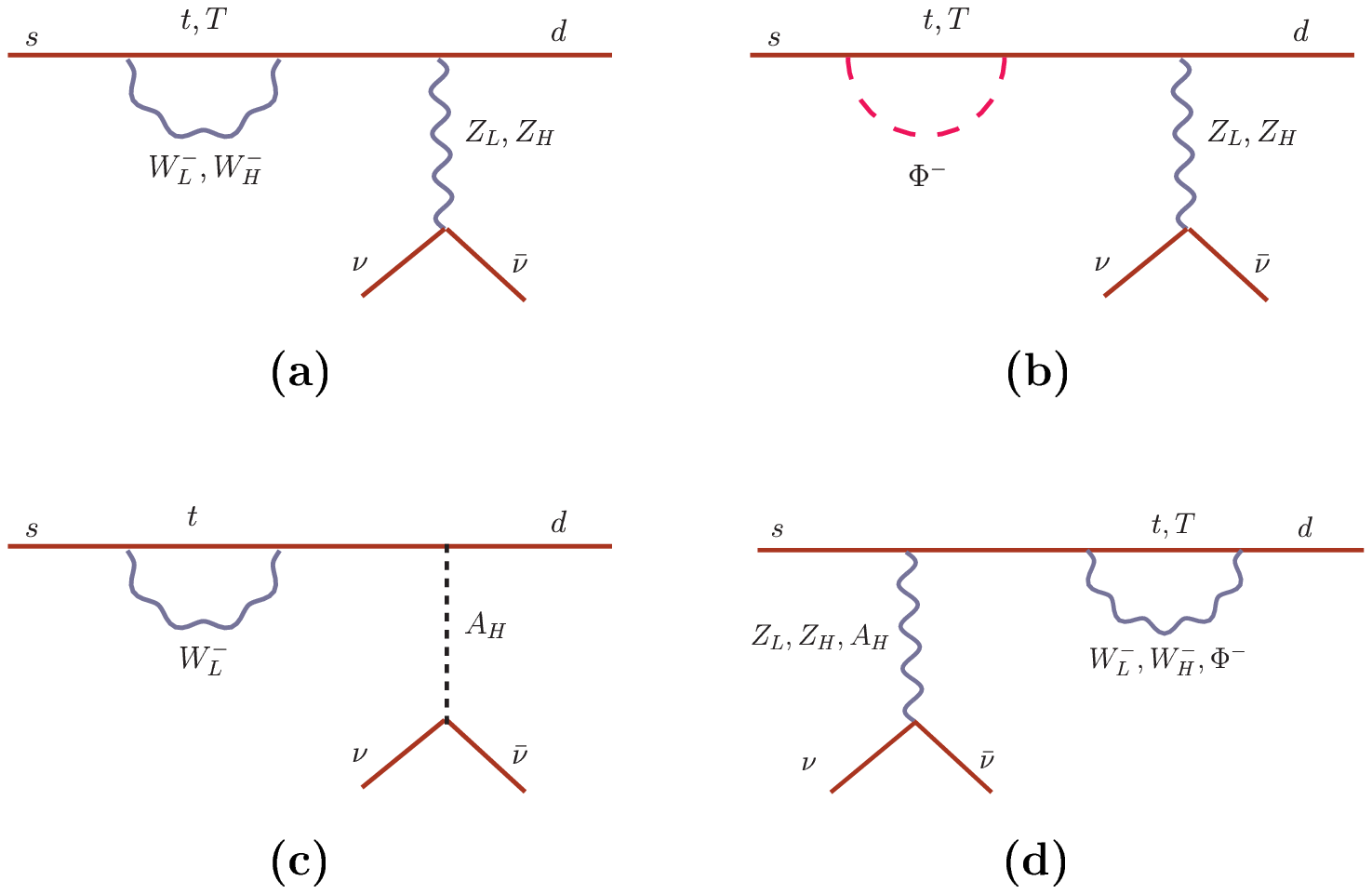,height=3in}
\caption{Self energy}
\label{dia:3}
\ecen
\efig

The contribution of the self energy diagram given in Figure
\ref{dia:3}(a) is :
\beqa
X^{SE}(t) &=& - \sum_{i,k,a}
 \left\{ \frac{8 \pi^2 M_{Z_L}^2
Cos^2\theta_w}{M_{Z_a}^2} g_{tsW}(i,k) g_{\nu \nu Z}(a)
g^L_{ddZ}(a)\right\}    \nonumber \\
&& 
\Bigg[
   - \frac{1}{16 \pi^2} \frac{m_i^2}{12 M_k^2}  
   + \int^1_0 dx \left( - 2 \left(1 - x \right) \right) F_1(C_{i,k}) 
   + \int^1_0 dx \left\{ 2 x + \frac{(1 - x)}{2} \right\} 
      \frac{F(C_{ik})}{M_k^2}       \nonumber \\
&& 
   + \int^1_0 dx \left\{ 2 x + \frac{(1 - x)}{2} \right\} 
      F_1(C_{ik}) \frac{C_{ik}^2}{M_k^2}
\Bigg]
\label{eq:sec3:4}
\eeqa
where $x$ is the Feynman parameter and $C_{ik}$ is the function of
Feynman parameter $x$ and masses $m_i$ and $M_k$ as given in
eqn(\ref{app:a:5}). 
The contribution of Box diagram in Figure \ref{dia:2} can be written
as :
\beqa
X^{Box} &=&
\sum_{i,k,l ~{\rm and}~ i=k=l\ne 1}\left\{ - 4 \pi^2 M_{Z_L}^2
Cos^2\theta_w  g_{tsW}(i,k) g_{tdW}(i,l) g_{\nu \ell W}(k) g_{\nu \ell
W}(l) \right\}     
\Bigg[ 4 F(m_i,M_k,M_l) 
  + \frac{m_i^2}{4 M_k^2 M_l^2} F(M_k,M_l)  \nonumber \\
&&  + \frac{m_i^4}{4 M_k^2 M_l^2} F(m_i,M_k,M_l)
  - m_i^2 \left( \frac{1}{M_k^2} +
    \frac{1}{M_l^2}\right)F(m_i,M_k,M_l)  
+ \frac{1}{16 \pi^2} \frac{m_i^2}{8 M_k^2 M_l^2} 
\Bigg] 
\label{eq:sec3:5}
\eeqa
In addition to above calculated diagrams there should be another self
energy diagram represented by Figure \ref{dia:3}(d). But this is
proportional to $\not p_d$ is $p_d$ is the external momenta of $d$
quark and hence this contribution vanishes when $m_d \to 0$. 


\section{Scalar Boson contributions \label{section:4}}

Little Higgs models also have doubly charged Higgs scalars but they
can't couple to SM fermions \cite{Han:2003wu,Schmaltz:2002wx} so they
won't give any contributions to $X^{LH}$ but the singly charged Higgs
($\Phi^\pm$) 
can give contributions in a manner similar to the contributions given
by charged vector bosons ($W^\pm_{L,H}$). The relevant Feynman
diagrams for the contribution of the charged scalars are given in
Figures \ref{dia:1}(c)(d),\ref{dia:3}(b). There won't be any extra
box diagram contribution in LH due to charged Higgs if neutrinos are
taken to be massless. 

\par We can write down the charged scalar contribution given in
eqn.(\ref{eq:sec2:2}) as :
\beq
X_{Z_{L,H}}(\Phi t, \Phi T) = X^\Phi(\Phi) + X^t(\Phi) + X^{SE}(\Phi)
\label{eq:sec4:1}
\eeq
where $X^\Phi(\Phi)$ and $X^t(\Phi)$ are the contributions of charged
Higgs penguins where $Z$ is emitted from charged Higgs line and top
quark line respectively. $X^{SL}$ is the self energy contribution. 
\beqa
X^\Phi(\Phi) &=& 
\sum_{i,a} \left\{ - \frac{8 \pi^2 M_{Z_L}^2 c_w^2}{g^2} 
\frac{(a - 2 s_w^2)}{c_w} \frac{1}{2 M_{Z_a}^2}
\frac{m_t^2}{v^2} \frac{v^2}{f^2} 
g^2_{dht}(i) g_{\nu \nu Z}(a) g_{hhz}(a)
\right\}            \nonumber     \\
&&{1 \over 4} 
\Bigg[
F_1(m_h) + m_i^2 F(m_i,m_h,m_h) + \frac{1}{32 \pi^2} 
\Bigg]
\label{eq:sec4:2}
\eeqa
where $m_h$ is the mass of the charged Higgs Boson and $\mu$ is the
$M_{W_L}$ mass scale. 
\beqa
X^t(\Phi) &=& \sum_{i,j,a}
\left\{ \frac{8 \pi^2 c_w^2 M_{Z_l}^2}{g^2} \frac{m_t^2}{v^2}
  \frac{v^2}{f^2} \frac{1}{M_{Z_a}^2} 
   g_{tsh}(i) g_{tdh}(j) g_{\nu \nu Z}(a) 
\right\}                     
\Bigg[ 
    g^L_{ttZ}(i,j,a) m_i m_j F(m_i,m_j,m_h)       \nonumber \\
&&
   - \frac{g^R_{ttZ}(i,j,a)}{2}
      \left\{ F(m_i,m_j) + m_h^2 F(m_i,m_j,m_h) 
         + {1 \over 4}\frac{1}{16 \pi^2}
      \right\}
\Bigg]
\label{eq:sec4:3}
\eeqa
\beqa
X^{SE}(\Phi) &=&
\sum_{i,a}
\left\{ \frac{8 \pi^2 c_w^2 M_{Z_l}^2}{g^2} 
      \frac{m_t^2}{v^2} \frac{v^2}{f^2} \frac{1}{M_{Z_a}^2}
      g_{dht}(i) g_{\nu \nu Z}(a) g^L_{ddZ}(a) 
\right\}  
    \int^1_0 dx \frac{1}{16 \pi^2} (1 - x) 
     log\left( \frac{\mu^2}{m_i^2 x + m_h^2 (1 - x)} \right)
\label{eq:sec4:4}
\eeqa


\section{Heavy Photon contribution \label{section:5}} 

For heavy photon ($A_H$), we don't have to consider the Higgs diagrams
because the Higgs couplings is already of order ${\cal O}(v/f)$, as
the mass of $A_H$ will come in the denominator \footnote{mass of Heavy
Photon is of order $f/v$}  so these diagrams would be higher order
diagrams in $(v^2/f^2)$. As the $WWA_H$ coupling is ${\cal
O}(v^2/f^2)$  so the penguin diagrams where $A_H$ is emitted from
$W^\pm$ would also be higher order diagram and hence need not be
considered. The coupling of the new vector type top quark ($T$) to
$W^\pm$ is proportional to $v/f$ and hence won't contribute to $A_H$
diagrams upto ${\cal O}(v^2/f^2)$. There won't be any contribution
from $W_H$ also because in case of $W_H$ diagrams the mass of heavy W
will come in denominator and hence makes the diagram to be of higher
order in $(v/f)$. 

\par Contribution of heavy photon diagrams can be written as :
\beq
X_{A_H}(t W_L) = X_{A_H}^{SE} + X_{A_H}^{peng}
\label{eq:sec5:1}
\eeq
\beqa
X_{A_H}^{peng} &=& 
\left\{- \frac{ (g'^2/g^2)}{M_{A_H}^2 s'^2 c'^2} 4 \pi^2
  M_{W_L}^2 \left( y_e - {4 \over 5} + \frac{c'^2}{2}\right)
\right\}  
\Bigg[ 
  g^L_{ttA}  \left( F_1(m_t) + M^2 F_1(m_t,M) 
\right.  \nonumber \\
&& \left.  
- \frac{2 m_t^2}{M^2} F_1(m_t)  + \frac{m_t^2 (m_t^2 - 2 M^2)}{M^2}
  F_1(m_t,M) \right)   
 + g^R_{ttA} m_t^2 \left( - {3 \over 2} F_1(m_t,M) + \frac{1}{2 M^2}
   F_1(m_t) \right) 
\Bigg]
\label{eq:sec5:2}    \\
X_{A_H}^{SE} &=& 
\left\{- \frac{ (g'^2/g^2)}{M_{A_H}^2 s'^2 c'^2} 4 \pi^2
  M_{W_L}^2 \left( y_e - {4 \over 5} + \frac{c'^2}{2}\right)
\right\} 
\int_0^1 dx
\Bigg[ - 2 (1 - x) F_1(C)  \nonumber \\
&& + \frac{(1 + 3 x)}{2 M^2}
     \left( F(C) + C^2 F_1(C) \right)
 - (1 - x) \frac{C^2}{32 \pi^2 M^2} 
\Bigg] 
\label{eq:sec5:3}
\eeqa
where in above expressions $m = m_t$ (mass of SM top quark) and $M =
M_{W_L}$ mass of SM W-boson and $C^2 = \left( x m^2 + (1 - x)
M^2\right)$ and $M_{A_H}$ is the mass of heavy photon.


\section{Numerical analysis and Discussion \label{section:6}} 

The Littlest Higgs model has a large spectrum of heavy particles other
than SM particles. In case of LH model there is a global SU(5)
symmetry which is broken at TeV range $\Lambda_s$ ($\Lambda_s = 4 \pi
f$) and in the process the scalar, both doublet and triplet acquire
vev's $v$ and $v'$. In our work we have used a model which has
considered model known as {\sl Littlest Higgs model} where we have
only a single light Higgs doublet but there are many variations of
this model possible which can extend this and have possibility of two
light Higgs doublets \cite{Skiba:2003yf}. 

\noindent Some of the universal features of all Little Higgs models
which are useful in phenomenology are : 
\begin{itemize}
\item{} Existence of heavy gauge bosons like $W^\pm_H$ and $Z_H$ which
are required to cancel $W$ and $Z$ loops.
\item{} A new heavy fermion which is required to cancel the SM top
quark divergence. 
\item{} Set of heavy scalars. This sector is heavily model dependent
and some variations of LH may have many singlets, doublets and
triplets. 
\end{itemize}
\bfig[t]
\bcen
\epsfig{file=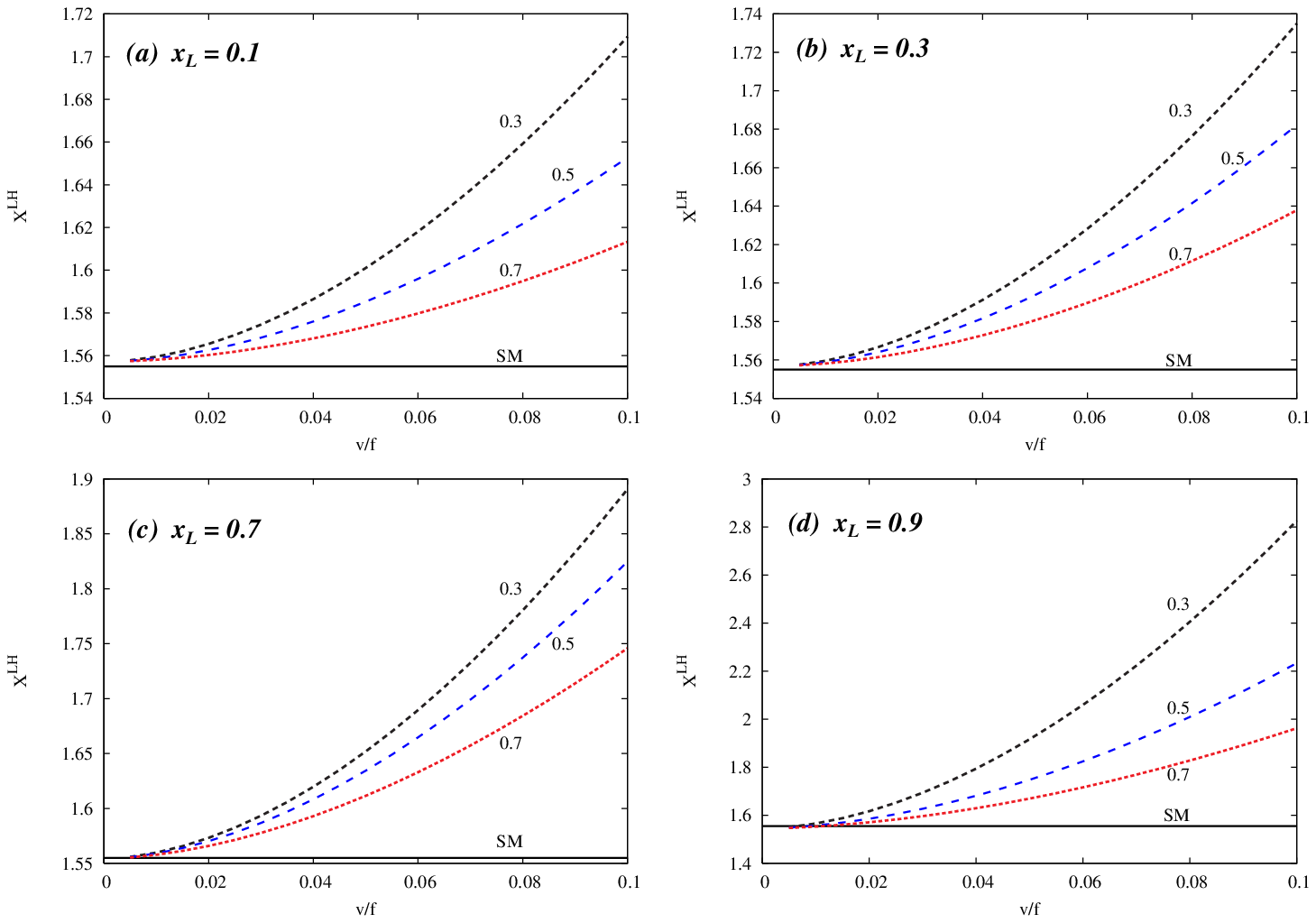,width=.95\textwidth}
\caption{Plot of $X^{LH}$ with $\frac{v}{f}$ for various values of
$s$. Different panels corresponds to different values of $x_L$. In
above plots we have used $s' = 0.4$.} 
\label{fig:result1}
\ecen
\efig

\noindent The Littlest Higgs model have following input parameters
above the SM one which can be parametrized as
\cite{Han:2003wu,Schmaltz:2002wx} : 
\begin{itemize}
\item[1.] $tan\theta = s/c = g_1/g_2$ ratio of new SU(2) coupling
constants. 
\item[2.] $tan\theta' = s/c = g_1'/g_2'$ ratio of new U(1) couplings.
\item[3.] $f$ : scale at which SU(5) global symmetry is broken.
\item[4.] $v'$ : vev of triplet Higgs. Triplet Higgs vev has a upper
bound given by $v' \le \frac{v^2}{4 f}$ where $v$ is the SM Higgs vev. 
\item[5.] $m_H$ : mass of SM Higgs boson.
\item[6.] $M_T$ : mass of the new vector type top quark. 
\end{itemize}

$M_T$ and $m_t$ (mass of SM top) together fixes two Yukawa couplings
$\lambda_1$ and $\lambda_2$ \cite{Han:2003wu}.

\par For our numerical analysis we will be going to use $s , s', v/f,
v'/f, m_H$ as input parameters. Regarding $M_T$ we will be going to
use another combination, which is the mixing parameter of SM top quark
and heavy vector like T quark, defined as $x_L =
\frac{\lambda_1^2}{\lambda_1^2 + \lambda_2^2}$ as the input parameter. 

Results of our numerical analysis are summarized in Figures
\ref{fig:result1} and \ref{fig:result2}. In these figures we have
plotted $X^{LH}$ as the function of various LH model parameters. 

In Figure\ref{fig:result1} we have plotted $X^{LH}$ as a function of
$v/f$ for various values of $s$. The four different panels in the plot
corresponds to four different $x_L$ values. The branching ratio of
$K_L \to \pi \nu \bar{\nu}$ is proportional to square of $X^{LH}$. As
we can see from the Fig.\ref{fig:result1} that the magnitude of
$X^{LH}$ in some region of LH parameter space can get a enhancement of
more than $100\%$ with respect to SM value which effectively means a
enhancement in branching ratio by a factor of four. 
We can also see
from this figure that LH model predicts substantial deviation from SM
results for higher $x_L$ values. The deviation from SM increases
for low $s$ values. 
\bfig[t]
\bcen
\epsfig{file=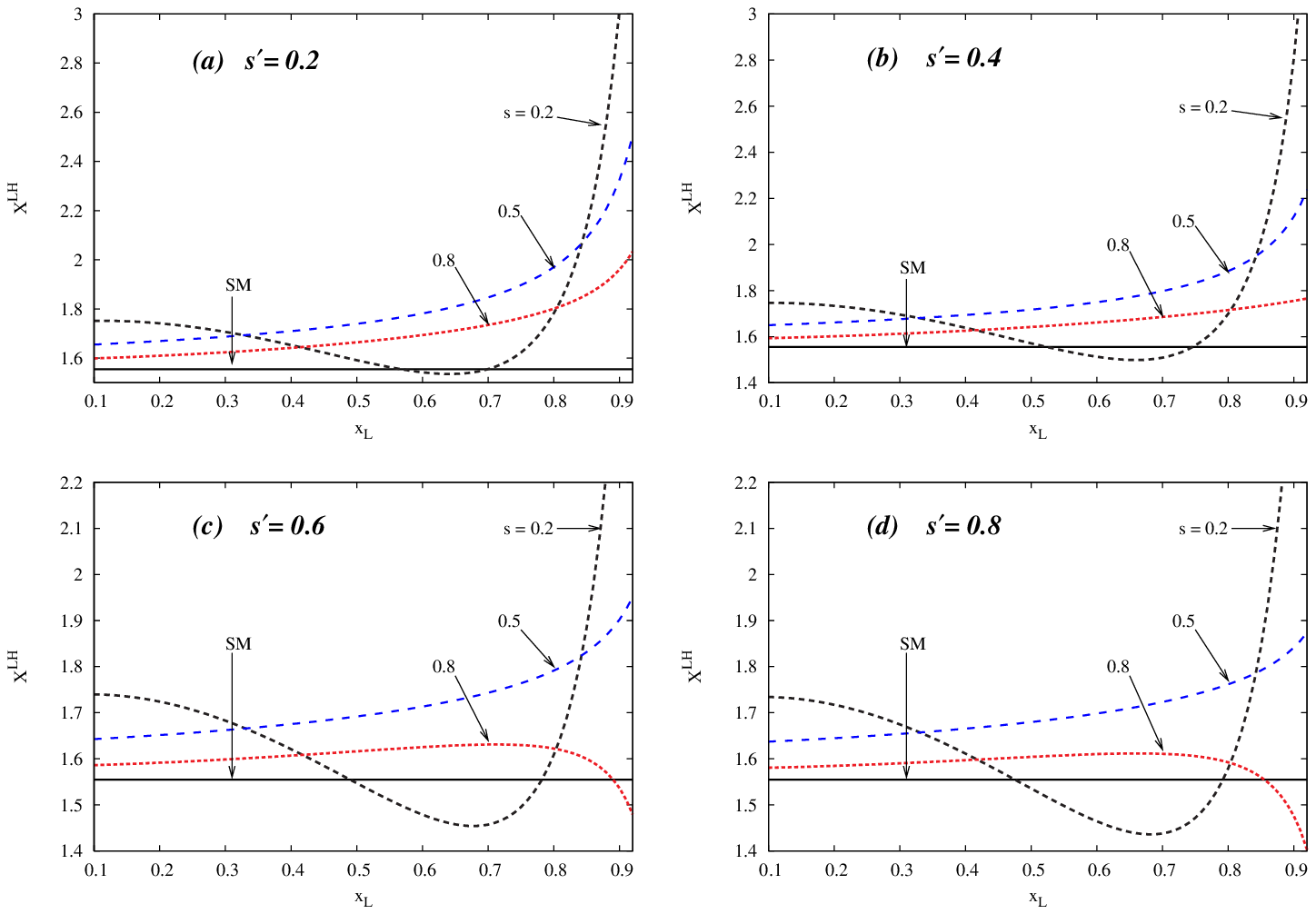,width=.95\textwidth}
\caption{Plot of $X^{LH}$ with $x_L$ for various values of
$s$. Different panels corresponds to different value of $s'$. In above
plots we have used $\frac{v}{f} = 0.1$}
\label{fig:result2}
\ecen
\efig
In Figure \ref{fig:result2} we have plotted $X^{LH}$ as a function of
$x_L$ for various values of $s$. Different panels corresponds to
different values of $s'$. For all the graphs in
Figure \ref{fig:result2} we have chosen the value $v/f = 0.1$. As we
can see from the figures that in certain region of parameter space we
can get substantial enhancements in the branching ratios. This figure
also emphasize that for higher $x_L$ values LH model can predict a
enhancement in the branching ratio of $K_L \to \pi \nu \bar{\nu}$ more
than 100\%. 

\par In this work we have confined ourselves to the {\sl Littlest
Higgs model} and have tried to investigate the effects extra top like
quark on $K_L \to \pi \nu \bar{\nu}$. The reason for choosing this
process was two fold firstly it is totally dominated by the top quark
exchange in SM hence relatively free from uncertainties secondly 
the branching ratio of this process ($K_L \to \pi \nu
\bar{\nu}$) scales with $m_t$ in SM. So in the sort of models which
predict the existence of a new quark whose behavior is similar to SM
top quark should effect this process the most. Heavier is the mass of
extra top like quark more increase it will predict to the branching
ratio of this process. Generically all the variations of Little Higgs
model predicts existence of a heavy top quark whose mass is in TeV
range and hence our qualitative results will remain the same in those
models also. Although this process is not yet observed but future
dedicated experiments would be able to observe this process 
\cite{Isidori:2003ij} and hence would be able to put constraints of
Little Higgs type models.  

\par Its is useful to discuss the constraints on parameters of LH
models imposed by various precision and decay measurements
\cite{Han:2003wu,Chen:2003fm,Hewett:2002px,Casalbuoni:2003ft,Csaki:2003si}.
In the Littlest Higgs model we don't have SU(2) {\sl 
custodial symmetry} \cite{Han:2003wu,Csaki:2003si}. This symmetry
protects relation of 
W, Z masses and  $\rho = 1$. The custodial SU(2) symmetry violating
corrections mainly arises from the heavy U(1) gauge bosons
\cite{Csaki:2003si}. Using this philosophy many variations of little
Higgs models which have approximate custodial SU(2) symmetry
\cite{Chang:2003un} have been constructed which can relax the
electroweak precision 
constraints on LH type models. These variations pushes the lower mass
bounds on heavy W, Z bosons and new top quark. 
As concluded by Chang \& Wacker \cite{Chang:2003un} that approximate
custodial symmetry can  
actually bring down the breaking scale (scale at which global symmetry
is spontaneously broken) upto 700 GeV. They also gave concluded that
precision electroweak measurements can push the lower bound on heavy W
and Z mass to around 2.5 TeV and mass of the top around 2 TeV. But as
pointed out above that we have chosen $K_L \to \pi \nu \bar{\nu}$
precisely for the reason that it is dominated by top in SM and hence
can have substantial modifications in these variations of LH which
pushes the mass of new top up. Low energy precision data on $(g -
2)_\mu$ of muon and atomic {\sl ``weak charge''} of cesium doesn't
impose any new constraint on model parameters \cite{Casalbuoni:2003ft} 

\par In Littlest Higgs model which we have considered Hewett
et.al. \cite{Hewett:2002px} noted that considering precision
electroweak measurement there exists a very small region of parameter
space where we can lower the bound of $f$ to be around TeV. But
considering Tevatron bounds also this bound on $f$ can be pushed to $f
\approx 3.5$ TeV.


\par We finally comment on the relationship of our estimate with a
recent paper by Buras et.al \cite{Buras:2004uu} who have studied this
process 
in relation to the observed anomalies in the decays of B into $\pi
\pi$ and $\pi K$ channels. The net conclusion of the paper was that in
order to explain the $\pi \pi, ~ \pi K$ anomaly one is led to an
effective value of $X$ slightly higher than the SM value but more
importantly with a phase of about $86^0$. The analysis is more or less
model independent but the likely area of the phase is in the squark
mass matrices. The LH model, which has been proposed as an alternative
to SUSY, however results in possible change in $|X|$ but no additional
phase. 


\begin{acknowledgments}
We would like to thank Heather Logan for useful discussions 
The work of SRC and NG is supported under the SERC scheme of
Department of Science \& Technology (DST), India under the project
no. SP/S2/K-20/99. SRC would like to thank School of Physics, Univ. of
Melbourne where this project was initiated. 
\end{acknowledgments}
\appendix
\section{Loop Functions : \label{appendix:a}}
\beqa
F(x) &=& \frac{x^2}{16 \pi^2} \Bigg[ 1 + Log\left( \frac{\mu^2}{x^2}
\right) \Bigg]           \label{ap:a:1}  \\
F_1(x) &=&  \frac{1}{16 \pi^2} Log\left( \frac{\mu^2}{x^2} \right) 
                         \label{ap:a:2} \\ 
F(x,y) &=& \frac{F(x) - F(y)}{x^2 - y^2}  \label{ap:a:3} \\
F(x,y,z) &=&  \frac{1}{(x^2 - y^2)} \Bigg[ F(x,z) - F(y,z) \Bigg] 
                                          \label{ap:a:4} \\ 
F_1(x,y) &=& - \frac{\left[ F(x) - F(y)\right]}{(x^2 - y^2)^2}
+ \frac{F_1(x)}{(x^2 - y^2)}    \label{ap:a:4a} \\ 
C_{ik}^2 &=&  x m_i^2 + M_k^2 (1 - x)       \label{app:a:5}
\eeqa

\section{Coupling constants : \label{appendix:b}} 

Various vertices which we have used in our calculations are defined in
Table \ref{table:1}.  
\begin{table}[h]
\bcen
\begin{tabular}{|| c  | c||} \hline  \hline
Particles  &  Vertex  \\ \hline  \hline 
$s t_i W_k$ & $\frac{i g}{\sqrt{2}} g_{tsW}(i,k) \gamma_\alpha P_L$ \\  
$d t_i W_k$ & $\frac{i g}{\sqrt{2}} g_{tsW}(i,k) \gamma_\alpha P_L$ \\
$\nu \bar{\nu} Z_a$ & $\frac{i g}{\sqrt{2}} g_{\nu \bar{\nu} Z}(a)
\gamma_\alpha P_L$  \\ 
$d d Z_a$   & $i g \left( g^L_{ddZ}(a) \gamma_\alpha P_L 
                  + g^R_{ddZ}(a) \gamma_\alpha P_R \right)$ \\ 
$\nu \ell W_k$  & $\frac{i g}{\sqrt{2}} g_{\nu \ell Z}(k)
\gamma_\alpha P_L$  \\  
$t_i t_j Z_a$  &  $i g \gamma_\alpha \left[ g^L_{ttZ}(i,j,a) P_L
  + g^R_{ttz}(i,j,a) P_R \right]$ \\ 
$s t_i \Phi^\pm$  &  $\frac{-i m_t}{\sqrt{2} v}  \frac{v}{f} P_L
g_{tsh}(i)$ \\
$d t_i \Phi^\pm$   &  $\frac{-i m_t}{\sqrt{2} v} \frac{v}{f} g_{dht}(i)
P_R $ \\ 
$d d A_H$  &  $\frac{i g'}{2 s' c'} \gamma_\alpha \left[ g^L_{ddA} P_L
+ g^R_{ddA} P_R \right]$ \\ 
$t t A_H$  &  $\frac{i g'}{2 s' c'} \gamma_\alpha \left[ g^L_{ttA} P_L
+ g^R_{ttA} P_R \right] $ \\ 
$\nu \bar{\nu} A_H $ & $\frac{i g'}{s' c'} \gamma_\alpha \left( y_e -
{4 \over 5} + {1 \over 2} c'^2 \right) P_L$ \\
$W_k(k) W_l(k) Z_a$ &  $ - i g g_{WWZ}(k,l,a) \left[ k_\lambda
g_{\chi \phi} - 2 k_\chi g_{\lambda \phi} + k_\phi g_{\chi \lambda}
\right] $ \\ 
$\Phi_k(k) \Phi_l(k) Z_a$ & $\frac{i g}{c_W} (a - 2 s_W^2) k_\alpha
g_{hhZ}(a)$, ~~  $a = 0$ for LH. \\ 
\hline \hline 
\end{tabular}
\caption{}
\label{table:1}
\ecen
\end{table}

\noindent where $y_e = - {2 \over 5}$ and $P_{L,R} = \frac{(1 \mp
\gamma_5)}{2}$ . The full expression of the coupling constants can be
read off from Han \etal \cite{Han:2003wu}.  




\begin{thebibliography}{99}

\bibitem{Arkani-Hamed:2001nc}
 N.~Arkani-Hamed, A.~G.~Cohen and H.~Georgi,
 Phys.\ Lett.\ B {\bf 513}, 232 (2001)
 [arXiv:hep-ph/0105239]; 
%
 N.~Arkani-Hamed, A.~G.~Cohen, T.~Gregoire and J.~G.~Wacker,
 JHEP {\bf 0208}, 020 (2002)
 [arXiv:hep-ph/0202089] ; 
%
 N.~Arkani-Hamed, A.~G.~Cohen, E.~Katz, A.~E.~Nelson, T.~Gregoire and J.~G.~Wacker,
 JHEP {\bf 0208}, 021 (2002)
 [arXiv:hep-ph/0206020] ;
%
 N.~Arkani-Hamed, A.~G.~Cohen, E.~Katz and A.~E.~Nelson,
 JHEP {\bf 0207}, 034 (2002)
 [arXiv:hep-ph/0206021] ; 
%
 I.~Low, W.~Skiba and D.~Smith,
 Phys.\ Rev.\ D {\bf 66}, 072001 (2002)
 [arXiv:hep-ph/0207243].


 \bibitem{Han:2003wu}
 T.~Han, H.~E.~Logan, B.~McElrath and L.~T.~Wang,
 Phys.\ Rev.\ D {\bf 67}, 095004 (2003)
 [arXiv:hep-ph/0301040].


\bibitem{Schmaltz:2002wx}
 M.~Schmaltz,
 Nucl.\ Phys.\ Proc.\ Suppl.\  {\bf 117}, 40 (2003)
 [arXiv:hep-ph/0210415] ; 
%
 H.~E.~Logan,
 arXiv:hep-ph/0307340 ; 
%
 H.~E.~Logan,
 arXiv:hep-ph/0310151.


\bibitem{Choudhury:2004bh}
S.~R.~Choudhury, N.~Gaur, A.~Goyal and N.~Mahajan,
Phys.\ Lett.\ {\bf B 601},164(2004)
[arXiv:hep-ph/0407050] ;
%
A.~J.~Buras, A.~Poschenrieder and S.~Uhlig,
arXiv:hep-ph/0410309.


\bibitem{Chen:2003fm}
 M.~C.~Chen and S.~Dawson,
 arXiv:hep-ph/0311032 ;
%
 C.~x.~Yue and W.~Wang,
 Nucl.\ Phys.\ B {\bf 683}, 48 (2004)
 [arXiv:hep-ph/0401214].
%
 W.~Kilian and J.~Reuter,
 arXiv:hep-ph/0311095.
%
 S.~Chang and H.~J.~He,
 Phys.\ Lett.\ B {\bf 586}, 95 (2004)
 [arXiv:hep-ph/0311177].


\bibitem{Huo:2003vd}
 W.~j.~Huo and S.~h.~Zhu,
 Phys.\ Rev.\ D {\bf 68}, 097301 (2003)
 [arXiv:hep-ph/0306029].


\bibitem{Han:2003gf}
 T.~Han, H.~E.~Logan, B.~McElrath and L.~T.~Wang,
 Phys.\ Lett.\ B {\bf 563}, 191 (2003)
 [arXiv:hep-ph/0302188].


\bibitem{Fleischer:2002ys}
R.~Fleischer,
Phys.\ Rept.\  {\bf 370}, 537 (2002)
[arXiv:hep-ph/0207108] ;
%
A.~J.~Buras, A.~Romanino and L.~Silvestrini,
Nucl.\ Phys.\ B {\bf 520}, 3 (1998)
[arXiv:hep-ph/9712398] ;
%
Y.~Grossman and Y.~Nir,
Phys.\ Lett.\ B {\bf 398}, 163 (1997)
[arXiv:hep-ph/9701313] ;
%
G.~Buchalla and A.~J.~Buras,
Phys.\ Rev.\ D {\bf 57}, 216 (1998)
[arXiv:hep-ph/9707243] ; 


\bibitem{Buchalla:1998ba}
G.~Buchalla and A.~J.~Buras,
Nucl.\ Phys.\ B {\bf 548}, 309 (1999)
[arXiv:hep-ph/9901288].


\bibitem{Isidori:2003ij}
G.~Isidori,
eConf {\bf C0304052}, WG304 (2003)
[arXiv:hep-ph/0307014].


\bibitem{Buras:2004uu}
A.~J.~Buras, F.~Schwab and S.~Uhlig,
arXiv:hep-ph/0405132 . 


\bibitem{Buras:2004ub}
A.~J.~Buras, R.~Fleischer, S.~Recksiegel and F.~Schwab,
arXiv:hep-ph/0402112 ;
%
A.~J.~Buras, R.~Fleischer, S.~Recksiegel and F.~Schwab,
Phys.\ Rev.\ Lett.\  {\bf 92}, 101804 (2004)
[arXiv:hep-ph/0312259] ; 


\bibitem{Deshpande:2004xc}
N.~G.~Deshpande, D.~K.~Ghosh and X.~G.~He,
arXiv:hep-ph/0407021.


\bibitem{He:2004it}
X.~G.~He and G.~Valencia,
arXiv:hep-ph/0404229.


\bibitem{Inami:1980fz}
T.~Inami and C.~S.~Lim,
Prog.\ Theor.\ Phys.\  {\bf 65}, 297 (1981)
[Erratum-ibid.\  {\bf 65}, 1772 (1981)].


\bibitem{Buchalla:1995vs}
G.~Buchalla, A.~J.~Buras and M.~E.~Lautenbacher,
Rev.\ Mod.\ Phys.\  {\bf 68}, 1125 (1996)
[arXiv:hep-ph/9512380].
%
A.~J.~Buras,
arXiv:hep-ph/0307203.


\bibitem{Skiba:2003yf}
W.~Skiba and J.~Terning,
Phys.\ Rev.\ D {\bf 68}, 075001 (2003)
[arXiv:hep-ph/0305302].


\bibitem{Hewett:2002px}
J.~L.~Hewett, F.~J.~Petriello and T.~G.~Rizzo,
JHEP {\bf 0310}, 062 (2003)
[arXiv:hep-ph/0211218].


\bibitem{Casalbuoni:2003ft}
R.~Casalbuoni, A.~Deandrea and M.~Oertel,
JHEP {\bf 0402}, 032 (2004)
[arXiv:hep-ph/0311038] ;
%
A.~Deandrea,
arXiv:hep-ph/0405120.


\bibitem{Chang:2003un}
 S.~Chang and J.~G.~Wacker,
 Phys.\ Rev.\ D {\bf 69}, 035002 (2004)
 [arXiv:hep-ph/0303001]; 
%
S.~Chang, 
 JHEP {\bf 0312}, 057 (2003)
 [arXiv:hep-ph/0306034] ; 


\bibitem{Csaki:2003si}
 C.~Csaki, J.~Hubisz, G.~D.~Kribs, P.~Meade and J.~Terning,
 Phys.\ Rev.\ D {\bf 68}, 035009 (2003)
 [arXiv:hep-ph/0303236].
%
 C.~Csaki, J.~Hubisz, G.~D.~Kribs, P.~Meade and J.~Terning,
 Phys.\ Rev.\ D {\bf 67}, 115002 (2003)
 [arXiv:hep-ph/0211124].

\end{thebibliography}
\end{document}